\documentclass[runningheads]{llncs}
\usepackage{graphicx}
\usepackage{amsmath,bm}
\usepackage{amsfonts}
\usepackage[utf8]{inputenc}
\usepackage[colorinlistoftodos]{todonotes}
\usepackage{cite}
\usepackage{amsbsy}
\usepackage{graphicx}
\usepackage{array}
\usepackage{url}
\usepackage{calc}
\usepackage{longtable}
\usepackage{caption}
\usepackage{hyperref}
\usepackage{svg}
\newcommand{\orcid}[1]{\href{https://orcid.org/#1}{\includegraphics[width=10pt]{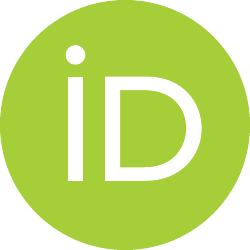}}}
\newcommand{\X}{\mathbf{X}}

\newcommand{\W}{\mathbf{W}}
\newcommand{\R}{\mathbf{R}}
\newcommand{\x}{\mathbf{x}}
\newcommand{\eps}{\bm \varepsilon}
\newcommand{\oo}{\bm{0}}
\newcommand{\z}{\mathbf{z}}
\newcommand{\tr}{\mathrm{tr}}
\newcommand{\Z}{\mathbf{Z}}
\newcommand{\I}{\mathbf{I}}
\renewcommand{\S}{\mathbf{S}}
\newcommand{\s}{\mathbf{s}}
\newcommand{\n}{\mathbf{n}}
\newcommand{\mmu}{\bm \mu}
\newcommand{\LLambda}{\bm \Lambda}
\newcommand{\SSigma}{\mathbf{\Sigma}}

\newcolumntype{G}{>{\centering\arraybackslash}m{4in+3\tabcolsep}}

\makeatletter
\newcommand{\printfnsymbol}[1]{%
  \textsuperscript{\@fnsymbol{#1}}%
}
\makeatother
\title{Functional Magnetic Resonance Imaging data augmentation through conditional ICA}
\DeclareUnicodeCharacter{2212}{-}
\begin{document}
 \author{Badr Tajini \orcid{0000-0001-8826-987X} \thanks{These authors contributed equally to this work},
 Hugo Richard \orcid{0000-0003-2229-7530} \printfnsymbol{1},
 Bertrand Thirion \orcid{0000-0001-5018-7895}}
 %index{Tajini, Badr}
 %index{Richard, Hugo}
 %index{Thirion, Bertrand}
%
\titlerunning{fMRI data augmentation through conditional ICA}
\authorrunning{Badr Tajini, Hugo Richard, Bertrand Thirion} 
\institute{Inria, CEA, Université Paris-Saclay, France}
\maketitle              
\begin{abstract}
  Advances in computational cognitive neuroimaging research are
  related to the availability of large amounts of labeled brain
  imaging data, but such data are scarce and expensive to generate.
  While powerful data generation mechanisms, such as Generative Adversarial Networks (GANs),  have been designed  in the last decade for computer vision, such improvements have not yet carried over to brain imaging.
  A likely reason is that GANs training is ill-suited to the noisy, high-dimensional and small-sample data available in functional neuroimaging.
  \thinspace In this paper, we introduce Conditional Independent Components Analysis (Conditional ICA): a fast functional Magnetic Resonance Imaging (fMRI) data augmentation technique, that leverages abundant resting-state data to create images by sampling from an ICA decomposition. We then propose a mechanism to condition the generator on classes observed with few samples.
  We first show that the generative mechanism is successful at
  synthesizing data indistinguishable from observations, and that it yields   gains in classification accuracy in brain decoding problems. In particular
  it outperforms GANs while being much easier to optimize and interpret. Lastly, Conditional ICA enhances classification accuracy in eight
  datasets without further parameters tuning.
\end{abstract}
\section{Introduction}
As a non-invasive brain imaging technique, task fMRI records brain
activity while participants are
performing specific cognitive tasks.
Univariate statistical methods, such as general linear models (GLMs)
\cite{friston1995analysis} have been successfully applied to
identifying the brain regions involved in specific tasks.
However such methods do not capture well correlations and interactions between brain-wide measurements.
By contrast, classifiers trained 
to \emph{decode} brain maps, i.e to discriminate between specific stimulus or task types~\cite{shirer_decoding_2012,varoquaux_how_2014,loula_decoding_2018}, take these correlations into account. 
The same framework is also popular for individual imaging-based diagnosis.

However, the large sample-complexity of these classifiers currently limits their accuracy.
To tackle this problem, data generation is an attractive approach, as
it could potentially compensate for the shortage of data.
Generative Adversarial Networks (GANs) are promising generative
models~\cite{goodfellow2014generative}.
However, GANs are ill-suited to the noisy, high-dimensional and
small-sample data available in functional neuroimaging. 
Furthermore the training of GANs is notoriously unstable and there are many hyper-parameters to tune.

In this work, we introduce Conditional ICA: a novel data augmentation technique using ICA together with conditioning mechanisms to generate surrogate brain imaging data and improve image classification performance.
Conditional ICA starts from a generative model of resting state data (unconditional model), that is fine-tuned into a conditional model that can generate task data. 
This way, the generative model for task data benefits from the abundant
resting state data and can be trained with few labeled samples.
We first show that the generative model of resting state data shipped in Conditional
ICA produces samples that neither linear nor non-linear classifiers are able to distinguish.
Then we benchmark Conditional ICA as a generative model of task data against
various augmentation methods including GANs and conditional GANs on their
ability to improve classification accuracy on a large task fMRI dataset. We find that Conditional ICA yields highest accuracy improvements.
Lastly, we show on 8 different datasets that the use of Conditional ICA results in systematic improvements in classification accuracy ranging from 1\% to 5\%.
\begin{figure}
\centerline{\includegraphics[width=1\textwidth]{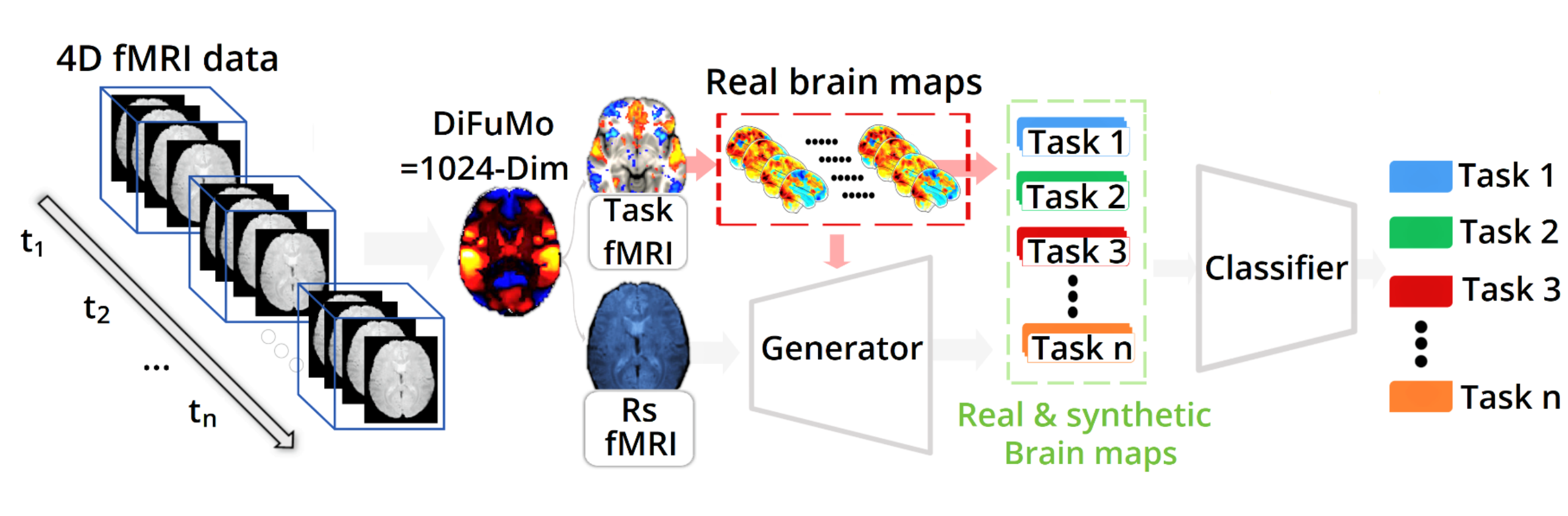}}
\caption{\textbf{Conditional ICA approach.} Our method aims to
  generate surrogate data from Task and Rest fMRI data by synthesizing
  statistical maps that qualitatively fit the distribution of the
  original maps. These can be used to improve the accuracy of
  machine learning models that identify contrasts from the
  corresponding brain activity patterns.}
\label{Fig0}
\end{figure}
\section{Methods}
\paragraph{Notations}
We write matrices as bold capital letters, vectors as small bold letters.
$\X^{\dagger}$ refers to the Moore-Penrose pseudo inverse of matrix $\X$,
, $\tr(\X)$ refers to the trace of matrix $\X$ and
$\I_k$ refers to the identity matrix in $\mathbb{R}^{k, k}$. $\oo_k$ refers to
the null vector in $\mathbb{R}^k$.

\paragraph{Spatial Dimension reduction.} 
The outline of the proposed approach is presented in Fig.\ref{Fig0}.
While brain maps are high-dimensional, they span a smaller space than that of
the voxel grid. 
For the sake of tractability, we reduce the dimension of the data by projecting the voxel values on the
high-resolution version of the Dictionaries of Functional Modes \emph{DiFuMo}
atlas \cite{dadi_fine-grain_2020}, i.e. with $p=1024$ components.
The choice of dimension reduction technique generally has an impact on the
results. However we consider this question to be out of the scope of the current study and leave this to future work.
\paragraph{Unconditional generative models (resting state data generation)}
Given a large-scale resting-state dataset $\X^{rest}$ in $\mathbb{R}^{p,n}$ where $n$ is the number of images (samples) and $p=1024$ the number of components in the atlas, let us consider how to learn its distribution.
Assuming a Gaussian distribution is standard in this setting, yet, as
shown later, it misses key distributional features.
Moreover, we consider a model that subsumes the distribution of any type of
fMRI data (task or rest): a linear mixture of $k \leq p$ independent temporal signals.
We therefore use temporal ICA to learn a dimension reduction and unmixing matrix
$\W^{rest} \in \mathbb{R}^{k, p}$ such that the $k$ sources i.e the $k$ components of
$\S^{rest} = \W^{rest} \X^{rest}$ are as
independent as possible.

A straightforward method to generate new rest data would be to
independently sample them from the distribution of the sources.
This is easy because such distribution has supposedly independent marginals.
We apply an invertible quantile transform $q^{rest}$ to the sources $S^{rest}$ so that
the distribution of $\z^{rest} =
q^{rest}(\s^{rest})$ has standardized Gaussian marginals. Since the distribution of
$\z^{rest}$ has independent marginals, it is given by $\mathcal{N}(\oo_k, I_k)$
from which we can easily sample.
As shown later, this approach fails: such samples are still separable
from actual rest data.

We hypothesize that this is because independence does not hold, and
thus a latent structure among the marginals of the source distribution has to be taken into account. Therefore we assume that the distribution of $\z^{rest}$ is given
by $\mathcal{N}(\oo_k, \LLambda^{rest})$ where $\LLambda^{rest}$ is a definite positive matrix.
$\LLambda^{rest}$ can easily be learned from a standard shrunk covariance
estimator: $\LLambda^{rest} = \SSigma^{rest} (1 - \alpha) + \alpha \tr(\SSigma^{rest}) \I_k$ where
$\alpha$ is given by the Ledoit-Wolf formula \cite{ledoit2004well} and
$\SSigma^{rest}$ is the empirical covariance of $\Z^{rest}$.

Our encoding model for rest data is therefore given by \\ 
$\Z^{rest} =
q^{rest}(\W^{rest} \X^{rest})$ and we assume that the
distribution of $\Z^{rest}$ is $\mathcal{N}(\oo_k, \LLambda_k)$.
The generative model is given by the pseudo inverse of the encoding model:
$\tilde{\X}^{rest} = (\W^{rest})^{\dagger} (q^{rest})^{-1}(\eps)$ where $\eps \sim
\mathcal{N}(\oo_k, \Lambda^{rest})$. 
\paragraph{Conditional generative models (generative model for task data)}
While resting state datasets have a large number of samples ($10^4 \sim 10^5$), task datasets   have a small number of samples ($10 \sim 10^2$). As a result, there are too few samples to learn high quality unmixing matrices. 
Therefore, using  the unmixing matrix $\W^{rest}$ learned from the resting state data, we rely on the following nonlinear generative model for brain maps in a certain class $c$:
\begin{equation}
  \x_c = (\W^{rest})^{\dagger} q^{-1}(\eps)
\end{equation}
with $\eps \sim \mathcal{N}(\mmu_c, \LLambda)$.

\begin{figure}
\centerline{\includegraphics[width=1\textwidth]{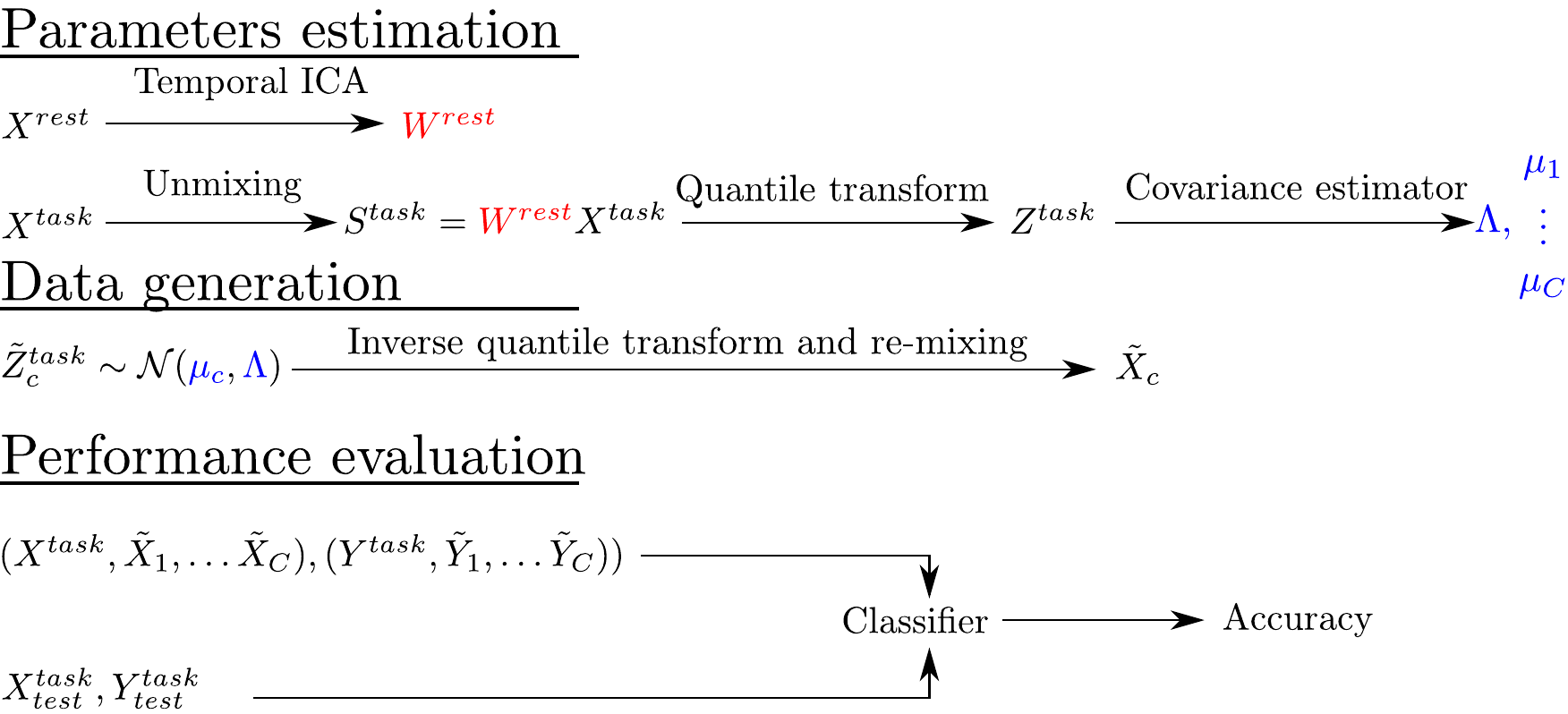}}
\caption{\textbf{Conditional ICA approach in depth.} 
The approach proceeds by learning a temporal ICA of rest data $\X^{rest} \in
\mathbb{R}^{p, n}$ , resulting in
independent sources and unmixing matrix $\W^{rest} \in \mathbb{R}^{k, p}$.
Applying the unmixing matrix to the task data, we obtain samples in the source
space $\S^{task} \in \mathbb{R}^{k, n}$.
Afterwards, we map $\S^{task}$ to a normal distribution, yielding $\Z^{task} \in
\mathbb{R}^{k, n}$. 
Then, we estimate the covariance $\LLambda \in \mathbb{R}^{k, k}$ (all classes are assumed to have the
same covariance) and the class-specific means $\mmu_1, \dots, \mmu_C \in \mathbb{R}^{k}$ according to Ledoit-Wolf's method.
For each class $c$, we can draw random samples $\tilde{\Z}^{task}_c \in
\mathbb{R}^{k, n_{\mathrm{fakes}}}$ from the
resulting multivariate Gaussian distribution $\mathcal{N}(\mmu_c, \LLambda)$ and
obtain fake data $\tilde{\X}_c  \in
\mathbb{R}^{p, n_{\mathrm{fakes}}}$
by applying the inverse quantile transform and re-mixing the data using the pseudo inverse of the unmixing matrix.
We append these synthetic data to the actual data to create our new augmented
dataset on which we train classifiers.}
\label{Fig11}
\end{figure}

In order to maximize the number of samples used to learn the parameters of the
model, we assume that the quantile transform $q$ and the latent covariance
$\LLambda$ do not depend on the class $c$. However, the mean $\mmu_c$, that can be learned efficiently using just a few tens of samples, depends on class $c$.
An overview of our generative method is shown in Fig.~\ref{Fig11}.

\section{Related work}
In image processing, data augmentation is part of standard toolboxes and
typically includes operations like cropping, rotation, translation.
On fMRI data these methods do not make much sense as brain images are not invariant to such transformations.
More advanced techniques~\cite{zhuang2019fmri} are based on generative models such as GANs or variational
auto-encoders~\cite{kingma2013auto}. Although GAN-based method are powerful they are slow and difficult to train~\cite{arjovsky_wasserstein_2017}. In
appendix Table~\ref{app:runningtime:tab}, we show that Conditional ICA is several orders of magnitude faster than GAN-based methods.

Our method is not an adversarial procedure. However it relates to other
powerful generative models such as variational
auto-encoders~\cite{kingma2013auto} with which it shares strong similarities.
Indeed the analog of the encoding function in the variational auto-encoder is
given by $e(\x) = \LLambda^{-\frac12}q(\W^{rest} \x)$ in our model and the analog to the decoding
function in the variational auto-encoder is given by $d(\z) =
(\W^{rest})^{\dagger}q^{-1}(\LLambda^{\frac12}\z)$ in our model. As in the variational auto-encoder, $e$ approximately maps the distribution of the data to a standardized Gaussian distribution,
while the reconstruction error defined by the difference in l2 norm
$\|d(e(\x)) - \x\|^2_2$ must remain small.

Another classical generative model related to ours is given by normalizing
flows~\cite{Kobyzev_2020}. When $\W^{rest}$ is squared (no dimension reduction in ICA), the decoding operator $d$ is invertible (its inverse is $e$) making our
model an instance of normalizing flows. 
A great property is thus the simplicity and reduced cost of data generation.
\paragraph{Software tools}
We use the FastICA algorithm~\cite{hyvarinen_fast_1999} to perform independent component analysis and use $k=900$ components. We use Nilearn~\cite{abraham_machine_2014} for fMRI
processing, Scikit-learn~\cite{pedregosa2011scikit},
Numpy~\cite{harris2020array} and Scipy~\cite{2020SciPy-NMeth} for data
manipulation and machine learning tools.
\paragraph{Code availability}
Our code is available on the following git repository:
\url{https://github.com/BTajini/augfmri}.
\section{Experiments}
\subsection{Dataset, data augmentation baselines and classifiers used}
The unmixing matrices are learned on the rest HCP
dataset~\cite{van2013wu} using 200 subjects.
These data were used after standard
preprocessing, including linear detrending, band-pass filtering
($[0.01, 0.1]Hz$) and standardization of the time courses.
The other 8 datasets~\cite{van2013wu, shafto2014cambridge,
  orfanos2017brainomics, pinel2019functional, pinel2007fast, pinel2013genetic,
  poldrack2016phenome, pinel2013genetic} are obtained from the Neurovault repository~\cite{gorgolewski2015neurovault}.
The classes used in each dataset correspond to the activation maps
related to the contrasts (such as ``face vs tools'')
present in the set of tasks of each dataset. Details are available in
appendix Table~\ref{app:dataset:tab}.

We consider 5 alternative
augmentation methods: \emph{ICA}, \emph{Covariance}, \emph{ICA + Covariance}, \emph{GANs} and \emph{CGANs}.
When no augmentation method is applied we use the label \emph{Original}.

The \emph{ICA} covariance method applies ICA to $\X^{task}$ to generate unmixing matrices $\W^{task}$ and
sources $\S^{task}=  \W^{task} \X^{task}$.
To generate a sample $\tilde{\x}_c$ from class $c$, we sample
independently from each source restricted to the samples of class $c$ yielding $\tilde{\s}^{task}_c$ and mix the data: $\tilde{\x}_c = (\W^{task})^{\dagger}
\tilde{\s}^{task}_c$.

The \emph{Covariance} method generates a new sample of
synthetic data in class $c$ by sampling from a Multivariate Gaussian
with mean $\mmu_c$ and covariance $\SSigma$, where $\mmu_c$ is the
class mean and $\SSigma$ is the covariance of centered task data
estimated using Ledoit-Wolf method.
In brief, it assumes normality of the data per class.

The \emph{ICA + Covariance} method combines the augmentation
methods \emph{ICA} and \emph{Covariance}: samples are drawn following
the ICA approach, but with some additive non-isotropic Gaussian noise.
As in \emph{ICA} we estimate $\W^{task}$ and $\S^{task}$ from
$\X^{task}$ via ICA.
Then we consider $\R_{task} = \X_{task} - \W_{task} \S_{task}$ and estimate the
covariance $\SSigma_R$ of $\R_{task}$ via LedoitWolf's method.
We then generate a data sample $\tilde{\x}_c$ from class $c$ as with ICA and add
Gaussian noise $\tilde{\n} \sim \mathcal{N}(0,\SSigma_R)$.
Samples are thus generated as $\tilde{\x}_c + \tilde{\n}$.

The \emph{GANs} (respectively \emph{CGANs}) method use a GANs (respectively CGANs) to generate data. The generator and discriminator have a mirrored architecture with 2 fully connected hidden layer of size (256 and 512).  The number of epochs, batch size, momentum and learning rate are set to 20k, 16, 0.9, 0.01 and we use the Leaky RELU activation function.

We evaluate the performance of augmentation methods through the use of classifiers: logistic regression (LogReg), linear
discriminant analysis with Ledoit-Wold estimate of covariance (LDA) perceptron
with two hidden layers (MLP) and random forrests (RF).
The hyper-parameters in each classifier are optimized through an internal 5-Fold
cross validation. We set the number of iterations in each classifier so that
convergence is reached. The exact specifications are available in appendix Table~\ref{app:classifiers:tab}.
\subsection{Distinguish fake from real HCP resting state data}
This experiment is meant to assess the effectiveness of the data
augmentation scheme in producing good samples.
Data augmentation methods are trained on 200 subjects taken from HCP rest fMRI
dataset which amounts to $960k$ samples (4800
per individual). Then synthetic
data corresponding to $200$ synthetic subjects are produced, yielding
$960k$ fake samples and various classifiers are trained to distinguish fake from
real data using 5-Fold cross validation. The cross-validated accuracy is shown
in Table~\ref{tab2}.
Interestingly, we observe a dissociation between linear models (LogReg
and LDA) that fail to discriminate between generated and actual data,
and non-linear models (MLP and RF) that can discriminate samples from
alternative augmentation methods.
By contrast, all classifiers are at chance when Conditional ICA is used.
\begin{table}
\begin{center}
\begin{tabular}{c|cccc}
\hline
Models & LDA  & LogReg & Random Forest &  MLP 
\\ \hline
ICA   & 0.493 & 0.500 & 0.672 &  0.697
\\
Covariance   & 0.473 & 0.461 & 0.610 &  0.626
\\
ICA + Covariance   & 0.509 & 0.495 & 0.685 &  0.706
\\
GANs~\cite{goodfellow2014generative}    & 0.501 & 0.498 & 0.592 &  0.607
\\
CGANs~\cite{mirza2014conditional}   & 0.498 &  0.493 & 0.579 & 0.604
\\\hline
\textbf{Conditional ICA}  & 0.503 & 0.489 & 0.512 &  0.523
\\\hline\hline
\end{tabular}
\end{center}
  \caption{\textbf{Distinguish fake from real HCP resting state data}
    We use HCP resting state data from $n=200$ subjects ($960k$ samples) and produce an equally
    large amount of fake data ($960k$ samples) using data augmentation methods.
    The table shows the 5-fold cross validated accuracy obtained with various
    classifiers. When Conditional ICA is used, all classifiers are at chance.
    }\label{tab2}
\end{table}
\subsection{Comparing augmentation methods based on classification accuracy on task
  HCP dataset}
In order to compare the different augmentation methods, we measure their 
relative benefit in the context of multi-class classification.
We use 787 subjects from the HCP task dataset that contains 23 classes and
randomly split the dataset into a train set that contains 100 subjects and a test set
that contains 687 subjects. In each split we train augmentation methods on the
train set to generate fake samples corresponding to $200$ subjects.  
These samples are then appended to the train set, resulting in an
augmented train set on which the classifiers are trained. Results, displayed in Table~\ref{tab3}, show that Conditional ICA always yields a higher accuracy  than when no augmentation method is applied. The gains are over 1\% on all classifiers tested. 
By contrast, ICA+Covariance and ICA lead to a decrease in accuracy
while the Covariance approach leads to non-significant
gains.
To further investigate the significance of differences between the proposed approach and other state-of-the-art methods, we perform a t-test for paired samples (see Table~\ref{app:significance} in appendix). Most notably, the proposed method performs significantly better than other data augmentation techniques. Given the large size of the HCP task data evaluation set, this significance test would demonstrate that the gains are robust.
Note that the random forest classifier is not used in this experiment as it leads to a much lower accuracy than other methods. For completeness, we display the results obtained with random forest in appendix Table~\ref{app:randomforrest}.

Visualization of fake examples produced by GANs, CGANs and Conditional ICA are available in appendix Figure~\ref{app:visualization:fig}.
\begin{table}[!t]
  \setlength{\tabcolsep}{0.23em}
\begin{center}
  \begin{tabular}{c|ccc||ccc||ccc}
    \hline
    Models &
      \multicolumn{3}{c}{LDA} &
      \multicolumn{3}{c}{LogReg} &
      \multicolumn{3}{c}{MLP} \\ \hline
    & Acc & Pre & Rec &  Acc & Pre & Rec & Acc & Pre & Rec \\
    \hline
    Original & 0.893 & 0.889 & 0.891 & 0.874 & 0.869 & 0.873 & 0.779 & 0.782 & 0.778 
    \\
    ICA   & 0.814 & 0.809 & 0.813 & 0.840 & 0.836 & 0.839 & 0.803  & 0.805 & 0.802 
    \\
    Covariance   & 0.895 & 0.894 & 0.895 & 0.876 & 0.877 & 0.875 & 0.819 & 0.823 & 0.820 
    \\
    ICA + Covariance   & 0.816 & 0.811 & 0.812 & 0.840 & 0.839 & 0.840 & 0.815 & 0.819 & 0.814  
    \\
    GANs~\cite{goodfellow2014generative} & 0.877 & 0.871 & 0.870 & 0.863 & 0.865 & 0.864  & 0.771 & 0.779 & 0.775
    \\
    CGANs~\cite{mirza2014conditional}  & 0.874 & 0.875 & 0.872 & 0.874 & 0.872 & 0.875 &  0.726 & 0.731 & 0.725 \\
    \hline
    \textbf{Conditional ICA} &  \textbf{0.901} & \textbf{0.903} & \textbf{0.905} & \textbf{0.890} & \textbf{0.888} & \textbf{0.890} & \textbf{0.832} & \textbf{0.835} & \textbf{0.831} \\
    \hline\hline
\end{tabular}
\end{center}
\caption{\textbf{Comparing augmentation methods based on classification accuracy on task
      HCP dataset} We compare augmentation methods based on the classification accuracy \textbf{(Acc)}, precision
    \textbf{(Pre)} and recall
    \textbf{(Rec)} obtained by 2 linear classifiers (LDA and LogReg) and one
    non-linear classifier trained on augmented datasets on HCP
  Task fMRI data. We report the mean accuracy across 5 splits.}\label{tab3}
\end{table}
\subsection{Gains in accuracy brought by conditional ICA on eight datasets.}
\begin{figure}[!b]
  \centering
  \includegraphics[width=\textwidth]{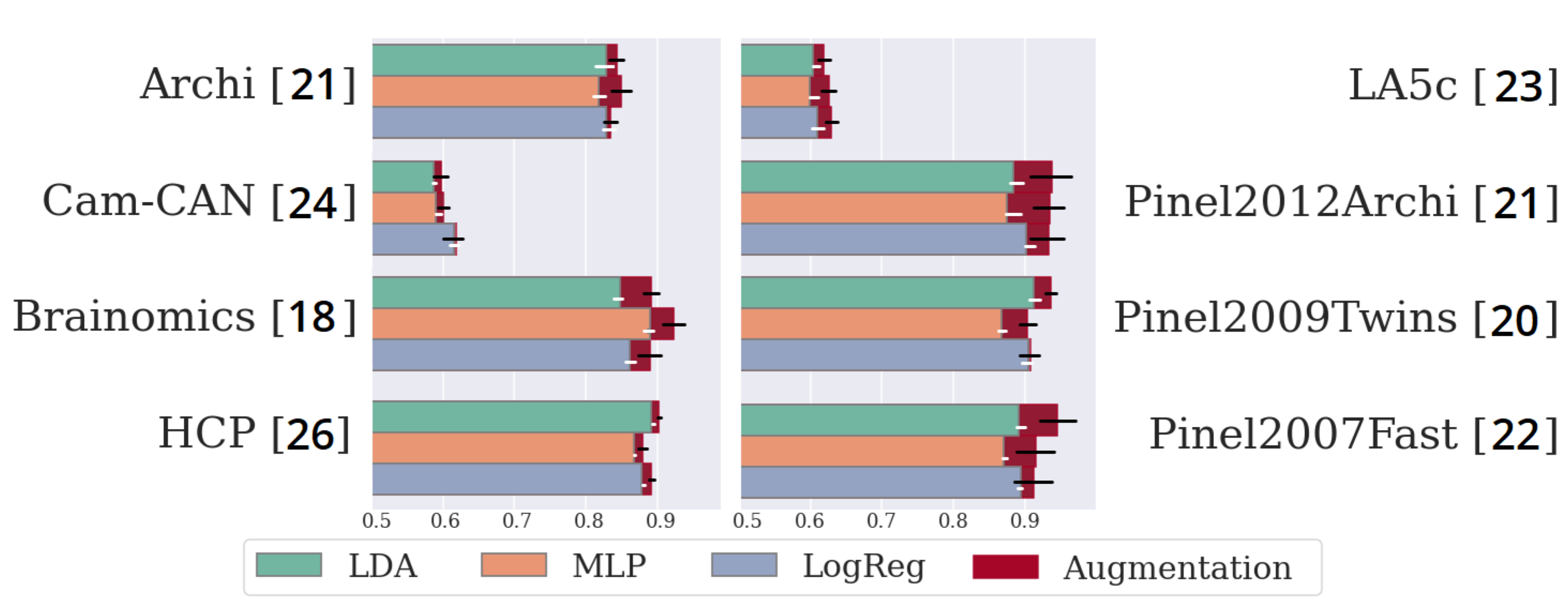}
\caption{\textbf{Accuracy of models for eight multi-contrast datasets.} Cross
  validated accuracy of two linear (LDA and LogReg) and one non-linear
  classifier (MLP) with or without using data augmentation.
  The improvement yielded by data augmentation is displayed in red.
  Black error bars indicate standard deviation across splits while white error bars indicate standard deviation across splits with no augmentation.}
\label{Fig4}
\end{figure}

In this experiment we assess the gains brought by Conditional ICA data
augmentation on eight different task fMRI dataset. The number of subjects,
classes and the size of the training and test sets differ between dataset and are reported
in appendix Table~\ref{app:classifiers:tab}. The rest of the experimental pipeline is exactly the same as with the HCP task dataset.
We report in Fig.~\ref{Fig4} the cross-validated accuracy of
classifiers with and without augmentation.
We notice that the effect of data augmentation is
consistent across datasets, classifiers and splits, with 1\% to 5\% net gains.
An additional experiment studying the sensitivity of Conditional ICA to the
number of components used is described in appendix Figure~\ref{app:sensitivity:fig}.
\section{Discussion \& Future work}
In this work we introduced Conditional ICA a fast generative model
for rest and task fMRI data. It produces samples that cannot be distinguished from true actual rest by 
linear as well as non-linear classifiers, showing that it captures higher-order statistics than naive ICA-based generators.
When Conditional ICA is used as a data augmentation method, it yields consistent
improvement in classification accuracy: on 8 tasks fMRI datasets, we observe
increase in accuracy between 1\% and 5\% depending on the dataset and the
classifier used.
Importantly, this performance was obtained without any fine-tuning of
the method, showing its reliability. One can also notice that our experiments cover datasets with different cardinalities, from tens to thousand, and different baseline prediction accuracy.
It is noteworthy that Conditional ICA is essentially a linear
generative model with pointwise non-linearity, which makes it cheap,
easy to instantiate on new data, and to introspect.

The speed and simplicity of Conditional ICA
as well as the systematic performance improvement it yields,
also make it a promising candidate for data augmentation in a wide range of
contexts. Future work may focus on its applicability to other decoding tasks
such as the diagnosis of Autism Spectrum Disorder
(ASD)~\cite{eslami2019asd,eslami2019auto,dvornek2017identifying} or
Attention-Deficit/Hyperactivity Disorder detection (ADHD)~\cite{mao2019spatio}. Other extensions of the present work concern the adaptation to
individual feature prediction (e.g. age) where
fMRI has shown some potential.
\section*{Acknowledgements}
This project has received funding from the European Union’s Horizon 2020 Framework Program for Research and Innovation under Grant Agreement No 945539 (Human Brain Project SGA3) and the KARAIB AI chair (ANR- 20-CHIA-0025-01).
\bibliographystyle{splncs04}
\bibliography{refs.bib}

\begin{thebibliography}{10}
\providecommand{\url}[1]{\texttt{#1}}
\providecommand{\urlprefix}{URL }
\providecommand{\doi}[1]{https://doi.org/#1}

\bibitem{abraham_machine_2014}
Abraham, A., Pedregosa, F., Eickenberg, M., Gervais, P., Mueller, A., Kossaifi,
  J., Gramfort, A., Thirion, B., Varoquaux, G.: Machine learning for
  neuroimaging with scikit-learn. Frontiers in Neuroinformatics  \textbf{8}
  (2014)

\bibitem{arjovsky_wasserstein_2017}
Arjovsky, M., Chintala, S., Bottou, L.: Wasserstein {GAN}. arXiv:1701.07875
  [cs, stat]  (2017)

\bibitem{dadi_fine-grain_2020}
Dadi, K., Varoquaux, G., Machlouzarides-Shalit, A., Gorgolewski, K.J.,
  Wassermann, D., Thirion, B., Mensch, A.: Fine-grain atlases of functional
  modes for fmri analysis. NeuroImage  \textbf{221},  117126 (2020)

\bibitem{dvornek2017identifying}
Dvornek, N.C., Ventola, P., Pelphrey, K.A., Duncan, J.S.: Identifying autism
  from resting-state fmri using long short-term memory networks. In:
  International Workshop on Machine Learning in Medical Imaging. pp. 362--370.
  Springer (2017)

\bibitem{eslami2019asd}
Eslami, T., Mirjalili, V., Fong, A., Laird, A.R., Saeed, F.: Asd-diagnet: a
  hybrid learning approach for detection of autism spectrum disorder using fmri
  data. Frontiers in neuroinformatics  \textbf{13}, ~70 (2019)

\bibitem{eslami2019auto}
Eslami, T., Saeed, F.: Auto-asd-network: a technique based on deep learning and
  support vector machines for diagnosing autism spectrum disorder using fmri
  data. In: Proceedings of the 10th ACM International Conference on
  Bioinformatics, Computational Biology and Health Informatics. pp. 646--651
  (2019)

\bibitem{friston1995analysis}
Friston, K.J., Holmes, A.P., Poline, J., Grasby, P., Williams, S., Frackowiak,
  R.S., Turner, R.: Analysis of fmri time-series revisited. Neuroimage
  \textbf{2}(1),  45--53 (1995)

\bibitem{goodfellow2014generative}
Goodfellow, I., Pouget-Abadie, J., Mirza, M., Xu, B., Warde-Farley, D., Ozair,
  S., Courville, A., Bengio, Y.: Generative adversarial nets. In: Advances in
  neural information processing systems. pp. 2672--2680 (2014)

\bibitem{gorgolewski2015neurovault}
Gorgolewski, K.J., Varoquaux, G., Rivera, G., Schwarz, Y., Ghosh, S.S., Maumet,
  C., Sochat, V.V., Nichols, T.E., Poldrack, R.A., Poline, J.B., et~al.:
  Neurovault. org: a web-based repository for collecting and sharing
  unthresholded statistical maps of the human brain. Frontiers in
  neuroinformatics  \textbf{9}, ~8 (2015)

\bibitem{harris2020array}
Harris, C.R., Millman, K.J., van~der Walt, S.J., Gommers, R., Virtanen, P.,
  Cournapeau, D., Wieser, E., Taylor, J., Berg, S., Smith, N.J., et~al.: Array
  programming with numpy. Nature  \textbf{585}(7825),  357--362 (2020)

\bibitem{hyvarinen_fast_1999}
Hyvärinen, A.: Fast and robust fixed-point algorithms for independent
  component analysis. Ieee Trans. Neural Netw  \textbf{10}(3),  626--634 (1999)

\bibitem{kingma2013auto}
Kingma, D.P., Welling, M.: Auto-encoding variational bayes. arXiv preprint
  arXiv:1312.6114  (2013)

\bibitem{Kobyzev_2020}
Kobyzev, I., Prince, S., Brubaker, M.: Normalizing flows: An introduction and
  review of current methods. IEEE Transactions on Pattern Analysis and Machine
  Intelligence p. 1–1 (2020). \doi{10.1109/tpami.2020.2992934},
  \url{http://dx.doi.org/10.1109/TPAMI.2020.2992934}

\bibitem{ledoit2004well}
Ledoit, O., Wolf, M.: A well-conditioned estimator for large-dimensional
  covariance matrices. Journal of multivariate analysis  \textbf{88}(2),
  365--411 (2004)

\bibitem{loula_decoding_2018}
Loula, J., Varoquaux, G., Thirion, B.: Decoding {fMRI} activity in the time
  domain improves classification performance. NeuroImage  \textbf{180},
  203--210 (2018)

\bibitem{mao2019spatio}
Mao, Z., Su, Y., Xu, G., Wang, X., Huang, Y., Yue, W., Sun, L., Xiong, N.:
  Spatio-temporal deep learning method for adhd fmri classification.
  Information Sciences  \textbf{499},  1--11 (2019)

\bibitem{mirza2014conditional}
Mirza, M., Osindero, S.: Conditional generative adversarial nets. arXiv
  preprint arXiv:1411.1784  (2014)

\bibitem{orfanos2017brainomics}
Orfanos, D.P., Michel, V., Schwartz, Y., Pinel, P., Moreno, A., Le~Bihan, D.,
  Frouin, V.: The brainomics/localizer database. Neuroimage  \textbf{144},
  309--314 (2017)

\bibitem{pedregosa2011scikit}
Pedregosa, F., Varoquaux, G., Gramfort, A., Michel, V., Thirion, B., Grisel,
  O., Blondel, M., Prettenhofer, P., Weiss, R., Dubourg, V., et~al.:
  Scikit-learn: {Machine} learning in {Python}. the Journal of machine Learning
  research  \textbf{12},  2825--2830 (2011)

\bibitem{pinel2013genetic}
Pinel, P., Dehaene, S.: Genetic and environmental contributions to brain
  activation during calculation. Neuroimage  \textbf{81},  306--316 (2013)

\bibitem{pinel2019functional}
Pinel, P., d’Arc, B.F., Dehaene, S., Bourgeron, T., Thirion, B., Le~Bihan,
  D., Poupon, C.: The functional database of the archi project: Potential and
  perspectives. NeuroImage  \textbf{197},  527--543 (2019)

\bibitem{pinel2007fast}
Pinel, P., Thirion, B., Meriaux, S., Jobert, A., Serres, J., Le~Bihan, D.,
  Poline, J.B., Dehaene, S.: Fast reproducible identification and large-scale
  databasing of individual functional cognitive networks. BMC neuroscience
  \textbf{8}(1), ~91 (2007)

\bibitem{poldrack2016phenome}
Poldrack, R.A., Congdon, E., Triplett, W., Gorgolewski, K., Karlsgodt, K.,
  Mumford, J., Sabb, F., Freimer, N., London, E., Cannon, T., et~al.: A
  phenome-wide examination of neural and cognitive function. Scientific data
  \textbf{3}(1),  1--12 (2016)

\bibitem{shafto2014cambridge}
Shafto, M.A., Tyler, L.K., Dixon, M., Taylor, J.R., Rowe, J.B., Cusack, R.,
  Calder, A.J., Marslen-Wilson, W.D., Duncan, J., Dalgleish, T., et~al.: The
  cambridge centre for ageing and neuroscience (cam-can) study protocol: a
  cross-sectional, lifespan, multidisciplinary examination of healthy cognitive
  ageing. BMC neurology  \textbf{14}(1), ~204 (2014)

\bibitem{shirer_decoding_2012}
Shirer, W.R., Ryali, S., Rykhlevskaia, E., Menon, V., Greicius, M.D.: Decoding
  {Subject}-{Driven} {Cognitive} {States} with {Whole}-{Brain} {Connectivity}
  {Patterns}. Cerebral Cortex (New York, NY)  \textbf{22}(1),  158--165 (2012)

\bibitem{van2013wu}
Van~Essen, D.C., Smith, S.M., Barch, D.M., Behrens, T.E., Yacoub, E., Ugurbil,
  K., Consortium, W.M.H., et~al.: The wu-minn human connectome project: an
  overview. Neuroimage  \textbf{80},  62--79 (2013)

\bibitem{varoquaux_how_2014}
Varoquaux, G., Thirion, B.: How machine learning is shaping cognitive
  neuroimaging. GigaScience  \textbf{3}, ~28 (2014)

\bibitem{2020SciPy-NMeth}
Virtanen, P., Gommers, R., Oliphant, T.E., Haberland, M., Reddy, T.,
  Cournapeau, D., Burovski, E., Peterson, P., Weckesser, W., Bright, J., {van
  der Walt}, S.J., Brett, M., Wilson, J., Millman, K.J., Mayorov, N., Nelson,
  A.R.J., Jones, E., Kern, R., Larson, E., Carey, C.J., Polat, {\.I}., Feng,
  Y., Moore, E.W., {VanderPlas}, J., Laxalde, D., Perktold, J., Cimrman, R.,
  Henriksen, I., Quintero, E.A., Harris, C.R., Archibald, A.M., Ribeiro, A.H.,
  Pedregosa, F., {van Mulbregt}, P., {SciPy 1.0 Contributors}: {{SciPy} 1.0:
  Fundamental Algorithms for Scientific Computing in Python}. Nature Methods
  \textbf{17},  261--272 (2020). \doi{10.1038/s41592-019-0686-2}

\bibitem{zhuang2019fmri}
Zhuang, P., Schwing, A.G., Koyejo, O.: Fmri data augmentation via synthesis.
  In: 2019 IEEE 16th International Symposium on Biomedical Imaging (ISBI 2019).
  pp. 1783--1787. IEEE (2019)

\end{thebibliography}
\clearpage
\appendix
\begin{table}
\begin{center}
\begin{tabular}{c|c|c|c}
\hline
Dataset & Subjects, Classes  & Train/Test & Neurovault collection 
\\ \hline
hcp~\cite{van2013wu}  & 787, 23 & 100/687  & collection 4337
\\
cam-can \cite{shafto2014cambridge}  & 605, 5 & 100/505  & collection 4342
\\
brainomics \cite{orfanos2017brainomics}  & 94, 19 & 50/44  &  collection 4341
\\
archi \cite{pinel2019functional}  & 78, 30 & 40/38  & collection 4339
\\
la5c \cite{poldrack2016phenome}  & 191, 24 & 100/91  & collection 4343
\\
pinel2012archi \cite{pinel2019functional} & 76, 10 & 40/36  & collection 1952
\\
pinel2009twins \cite{pinel2013genetic}  & 65, 12 & 35/30  & collection 1952
\\
pinel2007fast \cite{pinel2007fast} & 133, 10 & 70/63  & collection 1952
\\\hline\hline
\end{tabular}
\end{center}
\caption{\textbf{Datasets used in the experiments.} The table provides
  references to the datasets that were used for our experiments, with
  the number of subjects, the number of classes, the number of subjects in train
  and test set in each cross validation split and the collection number in Neurovault.}
  \label{app:dataset:tab}
\end{table}
\begin{center}
  \begin{longtable}{ p{.11\textwidth} | p{.3\textwidth} |p{.5\textwidth}}
\hline
  Methods & Optimizer & Hyper-parameters \\
  \hline
LogReg & L-BFGS \newline ($20~000$ iterations) & inverse $L_2$ regularization
strength \newline in $\{0.0001, 0.001, 0.01, 0.1, 1 \}$ \\
  \hline
LDA  & Least-squares solver & Estimation of covariance \newline using Ledoit-Wolf's
                              method \\
  \hline
  RF &  - &  Default parameters in sklearn \\
  \hline
MLP  & Adam \newline ($20~000$ iterations, \newline momentum: $0.9$, \newline
batch size: $32$, \newline learning
       rate: $0.0001$) & $ReLU$ activation function, fully connected
                         architecture with two hidden layers both of size $1024$, L2
                         penalty coefficient: $10^{-5}$ \\
\hline \hline
\caption{\textbf{Optimizers and hyper-parameters of classifiers} For each classifier, we give the optimization method used as well as the value of hyper-parameters.}\label{app:classifiers:tab} 
\end{longtable}
 
\end{center}
\begin{table}
\begin{center}
\begin{tabular}{c|c|c|c}
\hline
Models & LDA  & LogR & MLP
\\ \hline
Original & P $\leq$ 0.01 & P $\leq$ 0.05 & P $\leq$ 0.0001 \\
 ICA & P $\leq$ 0.0001 & P $\leq$ 0.001 & P $\leq$ 0.001 \\
Covariance & P $\leq$ 0.05 & P $\leq$ 0.01 & P $\leq$ 0.05 \\
ICA + Covariance & P $\leq$ 0.0001 & P $\leq$ 0.0001 & P $\leq$ 0.01 \\
GANs & P $\leq$ 0.001 & P $\leq$ 0.001 & P $\leq$ 0.0001 \\
CGANs & P $\leq$ 0.001 & P $\leq$ 0.01 & P $\leq$ 0.0001 \\
\hline\hline
\end{tabular}
\end{center}
\caption{\textbf{Statistical significance of the differences between Conditional ICA and other augmentation methods} P-values obtained after a t-test for paired samples is performed on the data used to produce Table~\ref{tab3}.}\label{app:significance}
\end{table}
\begin{table}
\begin{center}
\begin{tabular}{c|c|c|c|c|c|c|c}
\hline
Methods & Original & ICA & COV. & ICA + COV. & GANs & CGANs & \textbf{Cond. ICA}\\
\hline
Random Forest & 0.782 & 0.778 & 0.780 & 0.780 & 0.780 & 0.779 & \textbf{0.783}\\
\hline\hline
\end{tabular}
\end{center}
\caption{\textbf{Accuracy of Random Forrest} Mean accuracy obtained from data used for producing the Table~\ref{tab3} using a Random Forest classifier according to the chosen augmentation method.}\label{app:randomforrest}
\end{table}
\begin{table}
\begin{center}
\begin{tabular}{c|c}
\hline
Methods & Running time (secs)
\\ \hline
GANs  & 12948.2 ($\approx$ 3,60 hr)
\\
CGANs  & 11015.1 ($\approx$ 3,05 hr)
\\
\textbf{Conditional ICA}  & 62 s 
\\
\hline
\hline
\end{tabular}
\end{center}
\caption{\textbf{Running time.} We display the running time of three methods used
  to generate synthetic data. Conditional ICA is several orders of magnitude faster than
  GANs or CGANs. In practice the computational overhead induced by Conditional ICA is negligible.}\label{app:runningtime:tab}
\end{table}
\begin{figure}
  \centerline{\includegraphics[width=1\textwidth]{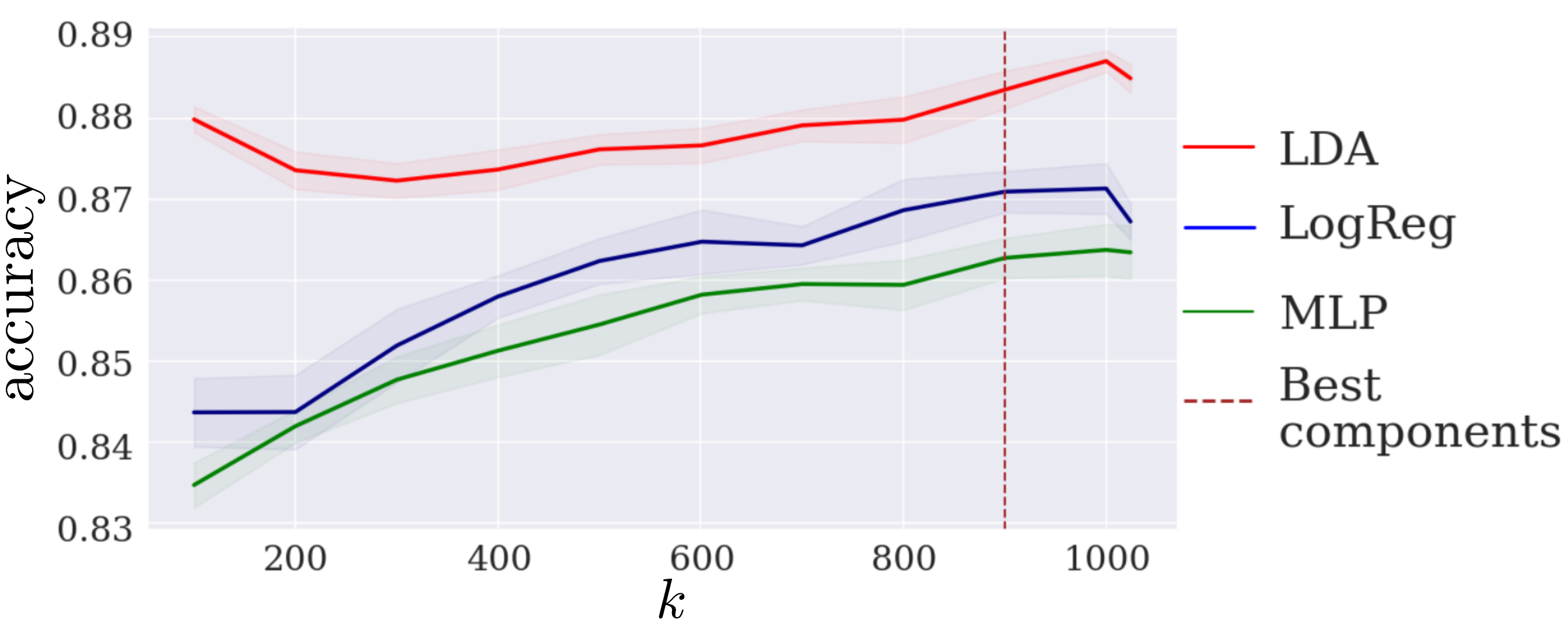}}
  \caption{\textbf{Accuracy of augmented discriminative models when
      varying $k$.} We use 100 train subjects from the HCP task dataset to train Conditional ICA with $k$ components and generate $200$ fake subjects.
    The classifiers are trained on the train and fake subjects and tested on the
    left-out 687 subjects. We repeat the procedure
    for various values of $k$ using 5 random splits per value and
    report the mean accuracy across splits as a function of $k$.
    The dotted line represents the number of components that has been
    used in our experiments ($k=900$).
  }
  \label{app:sensitivity:fig}
\end{figure}
\begin{figure}
  \centerline{\includegraphics[width=1\linewidth]{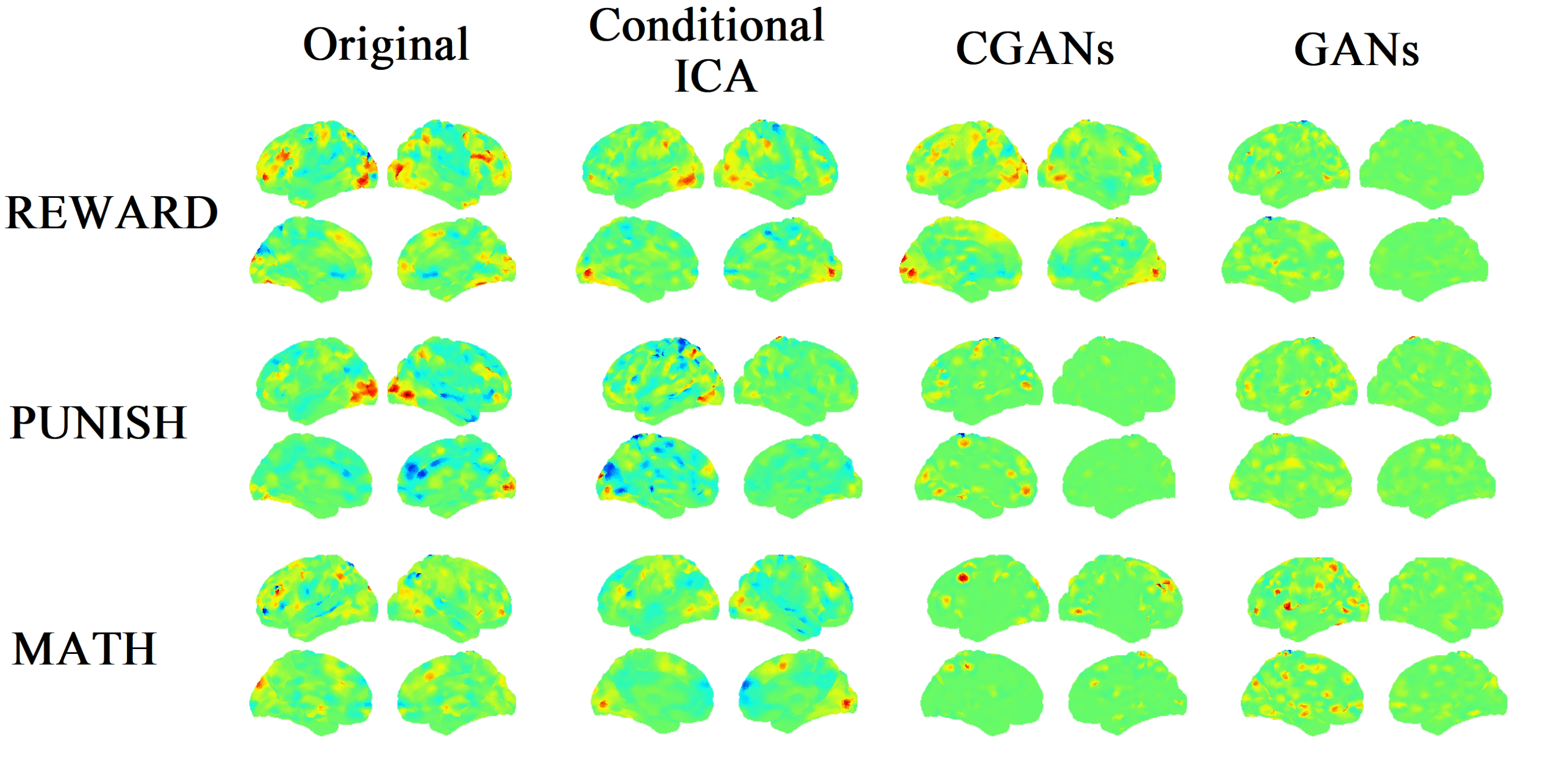}}
  \caption{\textbf{Data generation visualization.} Visualization of real
    (Original) and synthetic brain maps from three generation methods:
    Conditional ICA (Ours), CGANs and GANS. Three cognitive tasks are shown (reward, punish and math).
  }
  \label{app:visualization:fig}
\end{figure}
\end{document}